\documentclass{article}
\usepackage{spconf,amsmath,graphicx}
\usepackage[hidelinks]{hyperref}

\usepackage{times}
\usepackage{epsfig}
\usepackage{amssymb}
\usepackage{cite}

\usepackage{booktabs}
\let\svthefootnote\thefootnote
\newcommand\freefootnote[1]{%
  \let\thefootnote\relax%
  \footnotetext{#1}%
  \let\thefootnote\svthefootnote%
}
\PassOptionsToPackage{hyphens}{url}

\usepackage{color}

\def\revise#1{#1}

\title{SAVGBench: Benchmarking Spatially Aligned Audio-Video Generation}
\name{Kazuki Shimada$^{1,*}$, Christian Simon$^{2,*}$, Takashi Shibuya$^{1}$, Shusuke Takahashi$^{2}$, Yuki Mitsufuji$^{1,2}$}
\address{$^{1}$ Sony AI \hspace{0.1cm} $^{2}$ Sony Group Corporation}

\begin{document}
\ninept

\maketitle

\begin{abstract}
This work addresses the lack of multimodal generative models capable of producing high-quality videos with spatially aligned audio.
While recent advancements in generative models have been successful in video generation, they often overlook the spatial alignment between audio and visuals, which is essential for immersive experiences.
To tackle this problem, we establish a new research direction in benchmarking the Spatially Aligned Audio-Video Generation (SAVG) task.
We introduce a spatially aligned audio-visual dataset, whose audio and video data are curated based on whether sound events are onscreen or not.
We also propose a new alignment metric that aims to evaluate the spatial alignment between audio and video.
Then, using the dataset and metric, we benchmark two types of baseline methods: one is based on a joint audio-video generation model, and the other is a two-stage method that combines a video generation model and a video-to-audio generation model.
Our experimental results demonstrate that gaps exist between the baseline methods and the ground truth in terms of video and audio quality, as well as spatial alignment between the two modalities.
\freefootnote{$^*$ Equal contribution}
\end{abstract}


\section{Introduction}
\label{sec:intro}

Recently, generative models (e.g., diffusion models and transformer-based models) have shown remarkable achievement in generating high-quality videos~\cite{yang2025cogvideox,polyak2024movie,kong2024hunyuanvideo,wan2025wan,mao2024tavgbench,xing2024seeing,ruan2023mm}.
However, there are only a few models targeting multimodal generation, especially samples with audio-visual elements~\cite{mao2024tavgbench,xing2024seeing,ruan2023mm}.
Furthermore, videos generated by current state-of-the-art techniques fail to accurately represent real-world conditions, as they overlook spatial alignment between video and audio, which is necessary for creating immersive content.
The spatial component of audio-video not only enhances realism for experiential purposes but also provides a contextual understanding of the video (i.e., sound source directions), which can be used for many applications, such as virtual reality and world simulation.

This work establishes a new research direction in benchmarking the Spatially Aligned Audio-Video Generation (SAVG) task.
It aims to generate spatially synchronized multichannel audio and video, enhancing the overall user experience in multimedia applications.
While a few studies have tackled the task of generating multichannel audio from video~\cite{liu2025omniaudio,kim2025visage}, when considering the task of generating both multichannel audio and video (i.e., the SAVG task), a dataset and metrics tailored to the task are essential.
Regarding the dataset, considering the task of generating video as well, perspective video can be a natural choice, as it is widely used in video generation tasks.
The limited field of view results in both onscreen and offscreen sounds; however, including offscreen sound makes training and evaluation difficult.
Although several audio generation works employ spatial metrics using ground truth audio corresponding to the input video~\cite{liu2025omniaudio,kim2025visage,sun2025both}, we do not have ground truth audio corresponding to the \textit{generated} video in the SAVG task.

In this work, we propose SAVGBench, which comprises a well-curated audio-visual dataset and evaluation metrics for generated audio and video.
The audio-visual dataset contains perspective video and stereo audio, which are widely used formats in many types of media content.
It is derived from a previous audio-visual dataset~\cite{shimada2023starss23}, which consists of 360$^\circ$ video and its corresponding Ambisonics audio data, along with oracle position labels for sound events.
This 360$^\circ$ video data, Ambisonics audio data, and position labels provide omnidirectional coverage around the camera and microphone array, allowing us to track the position of sound events on and off the screen when we convert them into perspective video and stereo audio.
The video and audio data are curated based on whether sound events are onscreen or offscreen.
Its development and evaluation sets are released to the public\footnote{\url{https://zenodo.org/records/17139882}}.
For evaluation metrics, in addition to metrics that assess the video and audio quality and the temporal alignment, to evaluate the spatial alignment between video and audio, we propose a new metric based on detecting the positions of sounding objects in both modalities and measuring their alignment.
This metric relies on object detection~\cite{yolox2021} and sound event localization and detection (SELD)~\cite{adavanne2018sound,wilkins2023two}.
It is applicable even when both video and audio are generated, because no ground truth audio is required.

Using the SAVGBench dataset and metrics, we benchmark two types of baseline methods: joint and two-stage.
The joint method consists of an audio-visual diffusion model that focuses on joint audio-visual learning~\cite{ruan2023mm}.
It enables learning a joint distribution over both modalities~\cite{ruan2023mm}.
The two-stage method combines a video generation model and a video-to-audio generation model~\cite{cheng2025mmaudio}.
Note that we tested another two-stage method that combines an audio generation model and an audio-to-video generation model~\cite{yariv2024diverse}, but it did not work well in the dataset.
We evaluate the two baseline methods \revise{in an unconditional setup using the proposed dataset and metrics.}
The code is publicly available\footnote{\url{https://github.com/SonyResearch/SAVGBench}}.
We further conduct a subjective test on these results in terms of spatial audio-visual alignment.

In this work, our contributions are three-fold:
\begin{enumerate}
\item We present the SAVG task, which enhances multimodal generation by integrating spatially aligned audio and video.
\item We introduce the SAVGBench, which comprises a well-organized audio-visual dataset and a novel spatial audio-visual alignment metric utilizing an object detector and a SELD model.
\item We benchmark two types of baseline methods (i.e., a joint method and a two-stage method) in terms of spatial alignment, considering both objective and subjective perspectives.
\end{enumerate}

\section{Related Work}
\label{sec:related}

Recently, there has been growing interest in spatial audio generative models that aim to enhance immersive experiences~\cite{liu2025omniaudio,kim2025visage,sun2025both}. ViSAGe~\cite{kim2025visage} introduces a technique for generating first-order Ambisonics (FOA) audio conditioned on camera conditions, while OmniAudio~\cite{liu2025omniaudio} leverages 360$^\circ$ video data to produce spatial audio in the same format. Although both methods demonstrate promising results within their respective setups, they rely on specific input modalities, namely 360$^\circ$ video or perspective video with explicit camera orientation, which limits their direct applicability to our benchmark.

Another research interest is in audio spatialization with visual cues~\cite{gao20192,parida2022beyond,liu2024visually}.
Mono2Binaural~\cite{gao20192} presents a technique for synthesizing binaural audio from monaural recordings and the corresponding videos by modeling the dummy head recordings. While this approach shows promise in enhancing spatial audio realism, it focuses on generating binaural audio, considering the dummy head systems or head-related transfer functions.

\section{SAVGBench}
\label{sec:dataset}

\subsection{Dataset Overview}
\label{ssec:dataset-overview}

\begin{figure}[t]
    \centering
    \includegraphics[width=0.90\linewidth]{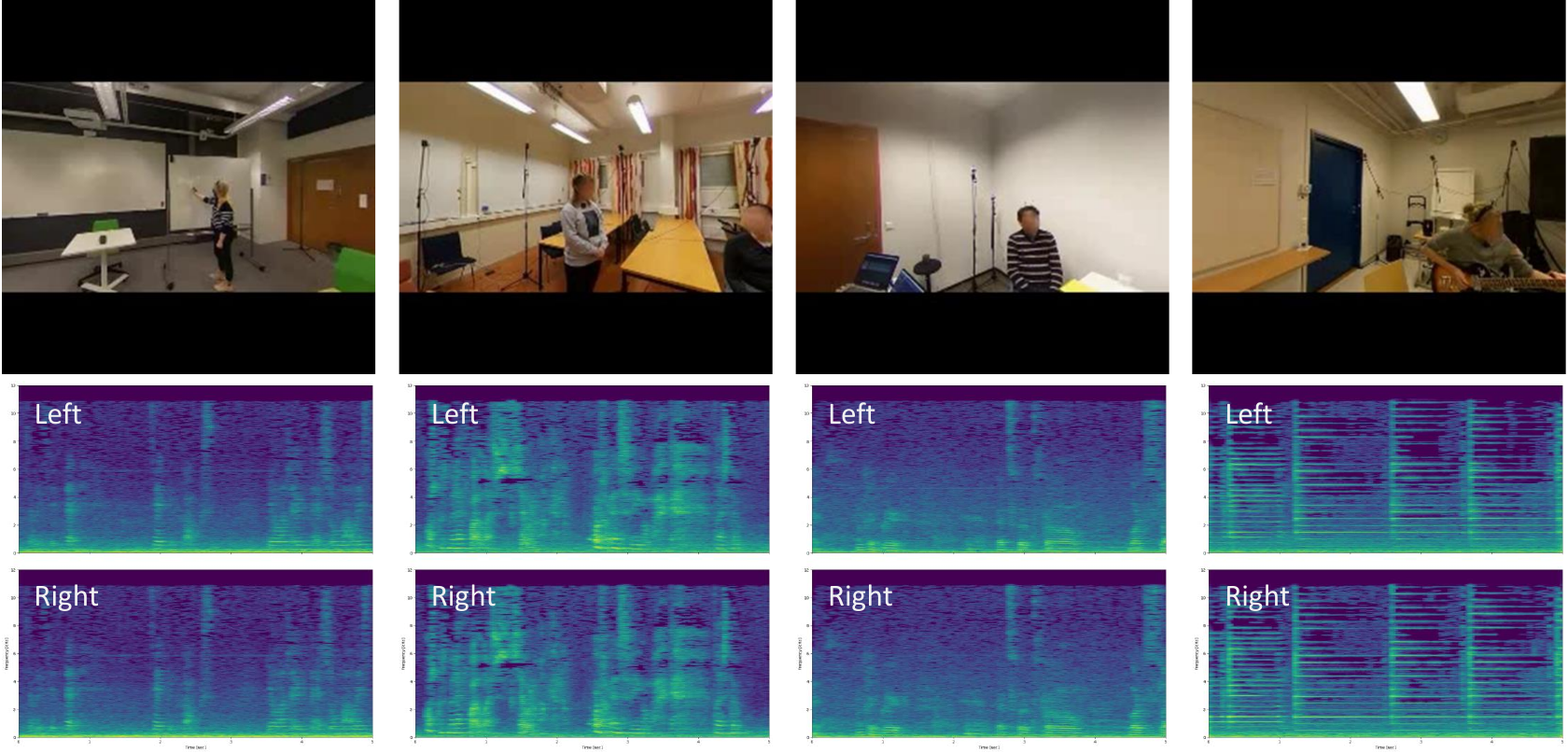}
    \vspace{-4mm}
    \caption{
    Examples of our proposed dataset.
    It features lectures, conversations, and playing musical instruments in indoor environments such as modern classrooms or meeting rooms.
    Perspective videos are square-shaped with padding.
    Stereo audio is displayed as a two-channel spectrogram.
    }
    \label{fig:example_spatial_svg24}
    \vspace{-5mm}
\end{figure}

We create an audio-visual dataset that contains stereo audio and perspective video data, derived from the STARSS23 dataset~\cite{shimada2023starss23}.
STARSS23 consists of sound scene recordings with various rooms and sound events, containing FOA audio data, the corresponding equirectangular video data, and spatiotemporal annotations, i.e., classes, activities, and positions of sound events.
We convert the STARSS23 data into stereo audio and perspective video data, tracking the position of sound events on and off the screen.
We curate audio and video data that contain only onscreen sound events for the proposed dataset.
It features humans and musical instruments in indoor environments, including speeches and instrument sounds.
The audio data are delivered as stereo audio with a 16 kHz sampling rate.
The video data uses a perspective view, ensuring that the video reflects sound events in the audio.
The video resolution is 256$\times$256 with padding.
The length and frame per second (fps) rate are set to 5 seconds and 4 fps, respectively.
We provide examples of the proposed dataset in Fig.~\ref{fig:example_spatial_svg24}.
It is split into the development set and the evaluation set.
The development set contains 5,031 videos, totaling about 7 hours.
The evaluation set serves as a target distribution to quantify the quality of generated videos and audio.
In both the development and evaluation sets, the ratio of speech to instrument sounds is maintained at about 2:1.
Note that this dataset has been used in a competition\footnote{\url{https://www.aicrowd.com/challenges/sounding-video-generation-svg-challenge-2024/problems/spatial-alignment-track}}.

\subsection{Data Construction}
\label{ssec:dataset-const}

We construct the proposed dataset as follows.
First, we extract 5-second videos every 0.5 seconds from the STARSS23 data.
Then, we convert the videos with the equirectangular view and FOA audio to videos with a perspective view and stereo audio, using a fixed viewing angle for the perspective view.
While we keep the vertical viewing angle at 0 degrees, we change the horizontal viewing angle by 10 degrees, sampling videos with a perspective view and stereo audio.
Finally, we curate videos that contain only onscreen speech and instrument sounds.
Note that the step sizes (i.e., 0.5 seconds and 10 degrees) are for the development set.
We use different step sizes for the evaluation set to keep the ratio of speech to instrument sounds at the same ratio as the development set.

We explain the details of data conversion, data curation, and other procedures.
\textbf{Data Conversion.}
Using a fixed viewing angle, we convert the equirectangular view and FOA audio to a perspective view and stereo audio.
\textbf{FOA $\to$ stereo:}
According to the viewing angle, \revise{we first rotate the FOA audio by applying a rotation matrix~\cite{mazzon2019first}.}
Then we convert the rotated FOA audio to the stereo with a simple translation~\cite{wilkins2023two}: $\texttt{left} = W + Y$ and $\texttt{right} = W - Y$, where $W$ is the omnidirectional signal of the FOA audio, and $Y$ is the first-order horizontal (left-right) component~\cite{wilkins2023two}.
\textbf{Equirectangular $\to$ perspective:}
We convert the equirectangular video to a perspective video with the same viewing angle as the audio, using a Python library\footnote{\url{https://github.com/sunset1995/py360convert}}.
We set the horizontal field of view to 100 degrees.
We also set the output height and width to 144 and 256, whose aspect ratio is 16:9, widely used in media content.
After the conversion, we add padding to make a video with 256$\times$256 resolution, as we use a pretrained super-resolution model of square ones in our baseline.

\textbf{Data Curation.}
We curate videos that contain only onscreen sound events.
Including only onscreen events facilitates the evaluation of the SAVG task.
During data conversion, we also convert position labels in the equirectangular video to position labels in a new perspective video.
Using the position labels in the perspective video, we investigate whether an event is onscreen or offscreen.
We also curate only events from speech and instrument classes, although the STARSS23 dataset contains other sound event classes.
The speech and instrument classes are stably detected by the object detector and SELD model, leading to a stable evaluation of spatial alignment.
In our preliminary experiments with the object detector, the person class was well detected, whereas other classes, such as cell phones or sinks, were not reliably detected in the 256$\times$256 resolution videos.
So, we considered using only human body-related classes, i.e., speech, clap, laugh, footstep, and instrument (as humans play instruments).
On the other hand, when we trained a SELD model with the human body-related classes, the SELD model did not perform well in detecting the clap, laugh, and footstep classes.
Finally, we use the speech and instrument classes to create our proposed dataset.
We use class labels for each event to determine whether an event belongs to one of these target classes.

\textbf{Other Procedures.}
In addition to data conversion and curation, several steps are performed to produce the final version of the proposed dataset. We remove videos with overlapping events to focus on single-source cases.
We set a threshold for the total length of sound events in a video to 80\% to ensure that the video contains sufficient sound events.
When the total length of sound events is less than 4 seconds in a 5-second video, we remove the video.
We apply a high-pass filter to all the audio data and amplify it by 38 dB to enhance its sound event signals, stabilizing the training of our baseline methods.
We exclude the video if amplified audio is clipping.

\subsection{Evaluation Metrics}
\label{ssec:metric}

We introduce a set of metrics to assess the quality of video and spatial audio generated from a model.
To evaluate the quality of audio-visual samples, we use Fréchet video distance (FVD)~\cite{Unterthiner2019FVDAN}, kernel video distance (KVD)~\cite{Unterthiner2019FVDAN}, and Fréchet audio distance (FAD)~\cite{kilgour2018fr} metrics to assess video and audio quality.
In addition, we also apply temporal alignment between visual and sound events adopted from~\cite{yariv2024diverse}.
To measure spatial alignment between an audio sample and the corresponding video sample, we propose a new metric, \textbf{Spatial AV-Align}, which ranges from zero to one, with higher values indicating better alignment. 

\begin{figure}[t]
    \centering
    \includegraphics[width=0.92\linewidth]{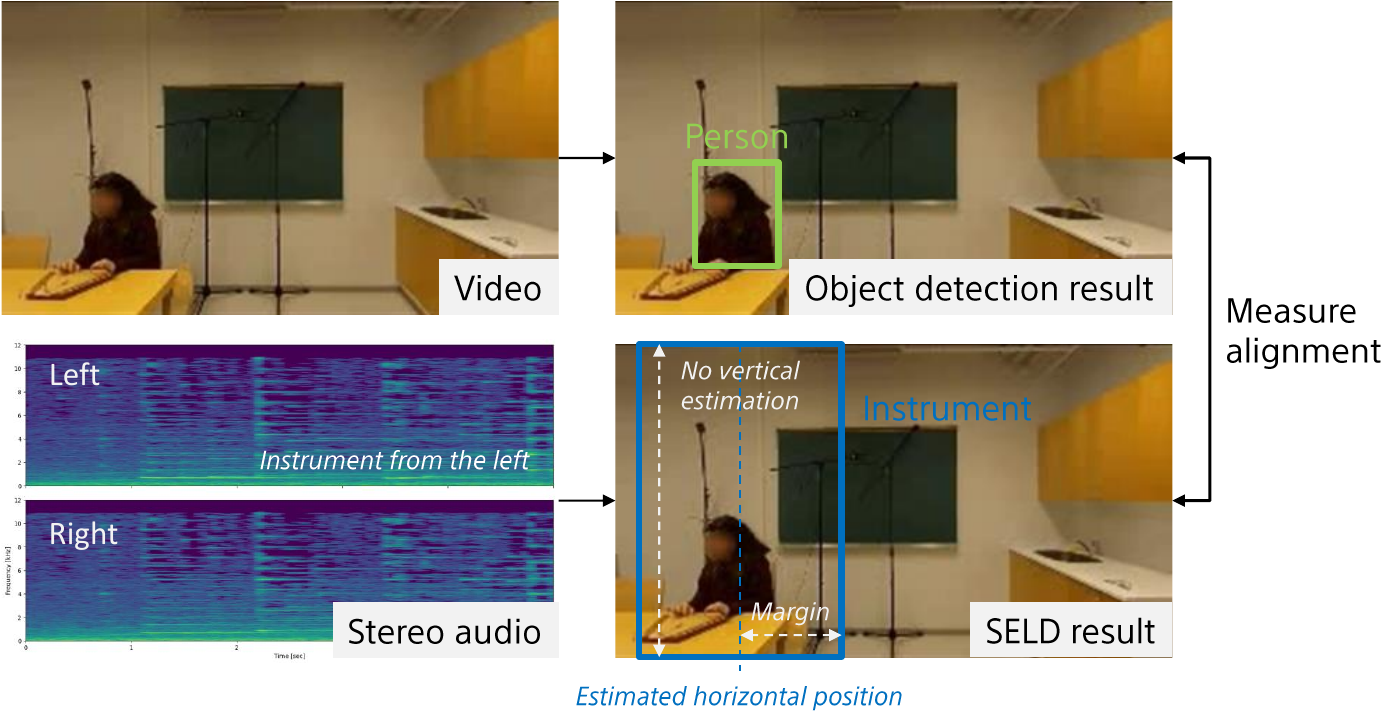}
    \vspace{-4mm}
    \caption{
    An illustration of the detected object (person class) and the detected sound event (instrument class) in the Spatial AV-Align metric.
    Padding is omitted for brevity.
    The green box indicates the detected object using the object detector.
    The blue box indicates the detected sound event using the SELD model.
    The SELD result has a margin around the estimated horizontal position.
    Its vertical range is set from top to bottom as it does not estimate a vertical position.
    }
    \label{fig:avalign_metric}
    \vspace{-5mm}
\end{figure}

The Spatial AV-Align metric evaluates spatial synchronization between sound events in audio and objects in video.
It relies on prior works on object detection~\cite{yolox2021} and SELD~\cite{adavanne2018sound,wilkins2023two}.
We use the widely known object detector YOLOX~\cite{yolox2021} to identify objects in video.
\revise{To detect and localize sound events in audio, we employ a stereo SELD model that processes stereo audio to estimate the activity and horizontal position of each class per frame.
Comprising convolution blocks, multi-head self-attention blocks, and fully connected layers, the model is trained on the development set of our proposed dataset using a combined loss of binary cross entropy and mean squared error.
Evaluation on the corresponding evaluation set shows that the model achieves an F-score exceeding 0.95 for all classes.}
See Fig.~\ref{fig:avalign_metric} for an illustration of a detected object (person class) and a detected sound event (instrument class) from each input.
\revise{The SELD result has a fixed margin around the estimated value in a horizontal position.
The vertical range of a SELD result is set from the top to the bottom.
Since stereo audio signals (left-right channels) do not contain vertical information, the model cannot handle vertical estimation.}
To evaluate this metric using stably detected objects or sound events, we focus on the person class in object detection and speech and instrument classes in SELD, as in the dataset section.

We explain the specific flow to compute the metric.
We begin by detecting candidate positions of sounding objects per frame within each modality separately.
Note that the fps for each modality is different: the object detector outputs at 4 fps, whereas the SELD model outputs at 10 fps.
Afterward, for a detected position in an audio frame, we verify if an object is also detected at the same position in the closest video frame to the audio frame.
Specifically, we determine if the SELD result overlaps with an object detection result.
If there is an overlap, it is considered as a true positive; otherwise, it is a false negative.
We do not verify if a detected object in the video aligns with the audio frames, as the dataset includes silent individuals and non-playing instruments.
Finally, we compute a recall metric as the alignment score, ranging from zero to one.
This alignment score is defined as $TP / (TP + FN)$, where $TP$ and $FN$ indicate the numbers of true positives and false negatives, respectively.

\section{Baseline Methods}
\label{sec:baseline}

\begin{table*}[!t]
    \centering
    \caption{
    Evaluation results of the baseline methods and ground truth on evaluation metrics.
    }
    \scalebox{1.00}{
        \begin{tabular}{lcccccc}
        \toprule
        \textbf{Method} & \textbf{FVD $\downarrow$} & \textbf{KVD $\downarrow$} & \textbf{FAD $\downarrow$} & \textbf{Temporal AV-Align $\uparrow$} & \textbf{Spatial AV-Align $\uparrow$} \\
        \midrule
         Ground Truth & 689.35 & 29.22 & 5.77 & 0.89 & 0.92 \\
        Joint Method: Stereo MM-Diff. & 1265.91 & 66.35 & 12.53 & 0.72 & 0.51 \\
        Two-stage Method: Video Diff. and Stereo MMAudio & 1386.53   & 71.82 & 12.00 & 0.78 & 0.35 \\
        \bottomrule
        \end{tabular}
    }
    \label{table:result_baseline}
    \vspace{-5mm}
\end{table*}

\subsection{Joint Method: Stereo MM-Diffusion}
\label{ssec:baseline-joint}
\vspace{-0.5mm}

In our benchmark, we develop a joint baseline method that extends MM-Diffusion~\cite{ruan2023mm} to the stereo case.
This strategy for audio-video alignment in diffusion models is by learning a joint distribution over audio and visual modalities. We employ a stereo channel extension of MM-Diffusion~\cite{ruan2023mm} as a baseline method. We use a structure similar to that of MM-Diffusion. The learning mechanism in MM-Diffusion implements joint learning between audio and video in the input space. Thus, the gaps between generated outputs between the two modalities are narrowed down.

In our implementation, MM-Diffusion consists of two separate branches for audio and video processing. In particular, the model encodes the source audio waveform using an audio encoder, yielding stereo audio with the size of $2 \times C \times T$, where $C$ and $T$ are the feature channel and corresponding time sequence. In the video branch, an encoder maps a video sequence of $F$ frames to dimensions of $F \times C \times H \times W$, where $H$ and $W$ are the height and width of a frame. The outputs from both the audio and video encoders are then integrated through a multi-modal attention module. This approach achieves stronger audio-video alignment than training each modality separately, particularly ensuring that the spatial origin and source of the audio are accurately matched. We follow the architectural setup of the original MM-Diffusion~\cite{ruan2023mm}, using 4 scales of MM-Blocks, each comprising 2 standard MM-Blocks, along with an additional downsample or upsample block.

As the model requires a huge amount of GPU memory, we only deal with videos of size $64 \times 64$ to fit a sample in a single GPU. In audio-visual generation with spatial alignment, precise positioning of objects is essential. An approach to obtaining object positions is to use an object detector. However, at the $64 \times 64$ resolution, the sounding objects are not visible, either to the object detector or even to the human eye. Therefore, a super-resolution model is required to upsample the video to $256 \times 256$ for better visibility of objects. The super-resolution model uses an identical architecture as in the guided diffusion model~\cite{dhariwal2021diffusion}.

As the model is designed unconditionally, the trained model could be used to generate a pair of audio-video samples. We observe that the DDPM~\cite{ho2020denoising}, along with the MM-Diffusion model, is slow to generate a sample. To expedite the testing process, we use DPM Solver~\cite{lu2022dpmsolver}. There is a trade-off in quality, but this issue is not significant as we only generate for a small resolution $64 \times 64$. For the super-resolution model, we still use DDPM to maintain the quality of the generated video.

\subsection{Two-stage Method: Video Diffusion and Stereo MMAudio}
\label{ssec:baseline-two}
\vspace{-0.5mm}

Below, we develop a two-stage baseline method that utilizes a portion of MM-Diffusion~\cite{ruan2023mm} as a video diffusion model and stereo-extended MMAudio~\cite{cheng2025mmaudio}.
This strategy involves training two separate models to handle video and stereo audio generation separately. The video diffusion model is trained unconditionally, using an architecture similar to Stereo MM-Diffusion, but without joint training alongside the audio modality. This architecture is proposed to ensure fair comparison between the joint training and two-stage methods. Since we follow the Stereo MM-Diffusion architecture, the pipeline does not include a VAE, which limits us to a low resolution of 64 $\times$ 64. To generate higher-resolution outputs, the video generation model employs the same super-resolution module used in Stereo MM-Diffusion.

In contrast to recent spatial audio generative models, e.g., OmniAudio~\cite{liu2025omniaudio} and ViSAGe~\cite{kim2025visage}, which focus on FOA audio with increased audio channel complexity, our work centers on a widely used stereo audio generation. To synthesize stereo sound from generated videos, we leverage MMAudio~\cite{cheng2025mmaudio}, a state-of-the-art model for video-to-audio generation. As the original MMAudio's VAE supports only mono audio, we extend the architecture to include two channels and apply a channel-wise decoding process, thereby enabling stereo audio generation. The video signal is processed with two visual models: Synchformer~\cite{iashin2024synchformer} for motion-audio alignment and CLIP~\cite{radford2021clip} for visual semantic features. The MMAudio structure is implemented using the Diffusion Transformer~\cite{peebles2023scalable} and Flow Matching~\cite{lipman2023flow} methods as originally proposed.

\vspace{-0.5mm}
\section{Experimental Evaluation}
\label{sec:exp}
\vspace{-0.5mm}

\vspace{-0.5mm}
\subsection{Experimental Setup}
\label{ssec:exp-setup}
\vspace{-0.5mm}

In our experiments, we use the development set of the proposed dataset to train the baseline methods, including the super-resolution model.
Note that the Stereo MM-Diffusion, video diffusion, and Stereo MMAudio models are trained from scratch.
On the other hand, the super-resolution model is initialized with a pretrained model from the guided diffusion model~\cite{dhariwal2021diffusion}, which is pretrained on ImageNet.
To finetune the super-resolution model, we extract all videos as frames and train the model on each frame.
Also, some components of Stereo MMAudio (i.e., VAE, Synchformer, and CLIP) are pretrained and kept frozen during training.
We set a batch size of 32 to train all models.
\revise{Further implementation details, such as the learning rate and number of iterations, are available in our GitHub repository.}
For evaluation, we use our proposed dataset’s evaluation set to assess model's performance.

\vspace{-0.5mm}
\subsection{Experimental Result}
\label{ssec:exp-result}
\vspace{-0.5mm}

We evaluate the quality and audio-visual alignment as shown in Table~\ref{table:result_baseline}.
We observe that the joint method achieves scores that are similar to or higher than those of the two-stage method. For distribution matching metrics (i.e., FVD, KVD, and FAD), the differences between the two approaches are minimal. Similarly, temporal AV-alignment scores are comparable. However, the joint method surpasses the two-stage method by a significant margin (approximately 0.15 points) in Spatial AV-Align scores. We hypothesize that this is because input conditions derived from generated videos are less plausible than those from ground truth videos and lack spatial information. In contrast, the joint method preserves spatial-temporal visual information during processing. Furthermore, compared to the Spatial AV-Align score on the evaluation set (i.e., ground truth), this result indicates potential for further improvement.

\vspace{-0.5mm}
\subsection{Subjective Test}
\label{ssec:exp-subjective}

\begin{table}[t]
    \centering
    \vspace{-2mm}
    \caption{Subjective test results of the baseline results and ground truth data in terms of spatial audio-visual alignment.}
    \scalebox{1.00}{
        \begin{tabular}{lc}
        \toprule
        \textbf{Method} & \textbf{MOS spatial audio-visual alignment $\uparrow$} \\
        \midrule
        Ground Truth & 4.06 \\
        Joint Method & 2.90 \\
        Two-stage Method & 2.44 \\
        \bottomrule
        \end{tabular}
    }
    \label{table:result_subjective}
    \vspace{-4.5mm}
\end{table}

\revise{We conduct the mean opinion score (MOS) test on spatial audio-visual alignment with 7 participants, who are recruited in person and have expertise in audio-visual technologies.
After passing a preliminary stereo listening test with a few samples using headphones, each participant then evaluates 45 randomly selected samples.
Participants are asked to evaluate the spatial alignment of each audio-video sample on a 5-point Likert scale while listening through headphones, based on the question: ``How would you rate the spatial alignment between the audio and video? (Focus on the direction of sounding sources and objects.)''}

In this test, we report a trend similar to that seen when evaluating spatial alignment among ground truth videos, the joint method, and the two-stage method, as shown in Table~\ref{table:result_subjective}. The joint method achieved a MOS of 2.90, whereas the two-stage method scored only 2.44 in this subjective listening evaluation. This result indicates that the joint method has better spatial alignment between visual and audio outputs compared to the two-stage method, following the trend in our Spatial AV-Align scores.

\section{Conclusion}
\label{sec:concl}

This paper presents a new benchmark for the Spatially Aligned Audio-Video Generation (SAVG) task.
The SAVGBench comprises a well-curated audio-visual dataset and evaluation metrics on generated audio and video.
The audio-visual dataset contains stereo audio and perspective video content that focuses on onscreen sound events, facilitating the training and evaluation of SAVG methods.
The Spatial AV-Align metric assesses spatial alignment between audio and video by utilizing an object detector and a sound event localization and detection (SELD) model.
It is applicable when both video and audio are generated, because ground truth audio is not required.
Using the dataset and metrics, we benchmark two types of baseline methods: one is based on a joint audio-video generation model, and the other is a two-stage method that combines a video generation model with a video-to-audio generation model.
Our experimental results indicate that gaps exist between the baseline methods and the ground truth in terms of video and audio quality, as well as spatial alignment between the two modalities.
\revise{This benchmark encourages future work to fill the gaps, especially in spatial audio-visual alignment.
Furthermore, other potential avenues include extending the benchmark to different indoor environments and more diverse sound events, and integrating text conditioning into the SAVG task.}

\newpage

\section{Acknowledgments}
\label{sec:ack}

\revise{We sincerely thank Ryosuke Sawata for his helpful feedback on this manuscript, Akio Hayakawa for his valuable advice on the MMAudio training and subjective test design, and Kengo Uchida for the constructive discussion about the evaluation metrics.}

\bibliographystyle{IEEEbib}
\bibliography{refs}

\end{document}